\documentclass[%
 aip,
 jcp,
 amsmath,amssymb,
 reprint,%
]{revtex4-2}

\usepackage[
mode=match,
separate-uncertainty=true,
propagate-math-font=true,
reset-math-version=false,
reset-text-family=false,
reset-text-series=false,
text-family-to-math=true,
text-series-to-math=true,
list-final-separator = {, and }
]{siunitx}
\DeclareSIUnit{\molar}{M}
\DeclareSIUnit{\inch}{in}
\DeclareSIUnit{\pixel}{px}

\usepackage{graphicx}
\graphicspath{{./figures/}{./supplemental-figures/}{./}}
\DeclareGraphicsExtensions{.pdf,.png}
\usepackage[colorlinks=true, citecolor=blue, linkcolor=blue]{hyperref}

\usepackage{dcolumn}
\usepackage{bm}

\usepackage[utf8]{inputenc}
\usepackage[T1]{fontenc}
\usepackage{textcomp}
\usepackage{mathptmx}
\usepackage{etoolbox}

\usepackage[textsize=tiny]{todonotes}
\usepackage{xargs}
\newcommandx{\vnm}[2][1=]{\todo[color=blue!30, #1]{vnm: #2}}
\newcommandx{\law}[2][1=]{\todo[color=red!30, #1]{law: #2}}
\newcommandx{\rfg}[2][1=]{\todo[color=green!30, #1]{rfg: #2}}
\newcommandx{\awp}[2][1=]{\todo[color=pink!30, #1]{awp: #2}}

\usepackage{url}
\usepackage{booktabs}

\makeatletter
\def\@email#1#2{%
 \endgroup
 \patchcmd{\titleblock@produce}
  {\frontmatter@RRAPformat}
  {\frontmatter@RRAPformat{\produce@RRAP{*#1\href{mailto:#2}{#2}}}\frontmatter@RRAPformat}
  {}{}%
}%
\makeatother

\begin{document}

\title{Electrostatic interactions drive the self-assembly of secondary
  coat-protein layers around wild-type MS2 particles}

\author{LaNell A. Williams}
\affiliation{Department of Physics, Harvard University, Cambridge, Massachusetts 02138, USA}
\affiliation{Center for the Physics of Biological Function, Princeton University, Princeton, New Jersey 08544, USA}

\author{Amelia W. Paine}
\affiliation{Harvard John A.\ Paulson School of Applied Engineering and
  Sciences, Harvard University, Cambridge, Massachusetts 02138, USA}

\author{Kellianne Kornick}
\affiliation{Department of Physics, Harvard University, Cambridge, Massachusetts 02138, USA}

\author{Rees F. Garmann}
\affiliation{Department of Chemistry and Biochemistry, San Diego State University, San Diego, California 92182 USA}
\affiliation{Viral Information Institute, San Diego State University, San Diego, California 92182 USA}

\author{Vinothan N. Manoharan}
\email{vnm@seas.harvard.edu}
\affiliation{Harvard John A.\ Paulson School of Applied Engineering and
  Sciences, Harvard University, Cambridge, Massachusetts 02138, USA}
\affiliation{Department of Physics, Harvard University, Cambridge, Massachusetts 02138, USA}

\date{\today}

\begin{abstract}
  The protein capsids of many icosahedral RNA viruses can self-assemble
  \textit{in vitro} in the absence of any host cell factors. For
  bacteriophage MS2 and related viruses, previous studies have focused
  on the role of specific RNA–protein interactions in driving the
  assembly process. However, the contribution of electrostatics remains
  unclear. We show that MS2 coat protein dimers can assemble into
  additional capsid layers around intact wild-type MS2 particles, even
  in the absence of free RNA. We characterize the size and morphology of
  the particles through transmission electron microscopy, gel
  electrophoresis, and dynamic light scattering, and we use binding and
  infectivity assays to determine how the maturation protein affects
  layer formation. Our results are consistent with a heterogeneous
  nucleation-and-growth assembly pathway, in which the nucleation
  barrier is controlled by the protein concentration, and electrostatic
  attraction between the positively charged coat proteins and negatively
  charged capsid drives growth. Interestingly, the secondary layers
  rarely fully encapsulate the wild-type particles, suggesting that
  elastic energy may limit their growth. These results highlight the
  importance of electrostatics in the assembly of protein shells, even
  in systems such as MS2 that are known for their strong, specific
  RNA-protein interactions.
\end{abstract}

\maketitle

\section{Introduction}

Self-assembly plays a crucial role in the replication of single-stranded
RNA (ssRNA) viruses. Cells infected by such viruses produce many copies
of the coat protein, which must then assemble around the viral RNA to
form the protective capsid.\cite{caspar1962physical} Remarkably, for
many simple RNA viruses, this assembly process occurs \textit{in vitro},
outside of the host cell, yielding virus-like particles (VLPs) with the
same structure as wild-type
particles.\cite{fraenkel-conrat_reconstitution_1955,
  bancroft_formation_1967, hiebert_assembly_1968,
  sugiyama1967ribonucleoprotein, beckett1988roles, peabody1993rna} For
bacteriophage MS2, a model system for making and studying VLPs, 180 coat
proteins assemble \textit{in vitro} around the viral RNA to form
icosahedral capsids in the VLP case. Although MS2 VLPs do not include
the maturation protein required for infection, they have the same
symmetry (triangulation number $T=3$) and size (about
\SI{30}{\nano\meter}) as the native
virus.\cite{sugiyama1967ribonucleoprotein} These \textit{in vitro}
results suggest that virus self-assembly can be understood as a physical
process, driven by interactions among the coat proteins and RNA.

A central question is what interactions drive this process. Previous
studies on MS2 and related bacteriophages have focused on the role of
the strong, sequence-specific interaction between the translational
operator region (TR loop) of the RNA and the coat
protein.\cite{witherell_specific_1991} Yet assembly of VLPs around RNA
molecules without this region can still occur, albeit at a higher
protein concentration or with reduced yield.\cite{beckett1988roles,
  witherell_specific_1991} These results point to the importance of
other interactions in driving assembly. As Beckett and colleagues noted
in their study of R17,\cite{beckett1988roles} a close relative of MS2,
``\dots the interaction of coat protein with non-specific sequences is
at least as important as its interaction with the specific translational
operator sequence.''

Arguably, the most important non-sequence-specific interaction is
electrostatic attraction, arising from the dissociable surface groups on
the RNA and coat protein and mediated by the ionic conditions of the
surrounding solution.\cite{zandi_virus_2020,
  losdorfer_bozic_varieties_2017} Electrostatic interactions are known
to drive the assembly of ssRNA plant viruses such as cowpea chlorotic
mottle virus (CCMV).\cite{garmann2014assembly} In MS2, however, the RNA
has both a negatively charged phosphate backbone as well as secondary
structures that interact specifically with coat
protein,\cite{peabody1993rna, stockley2007simple, valegard1997three}
making it difficult to disentangle the effects of electrostatics from
those of specific interactions. One way to probe the role of
electrostatics on assembly is to change the assembly template. For
example, replacing the native RNA with a charged synthetic
homopolymer\cite{hohn_fr_1969} or protein\cite{glasgow2012osmolyte}
would allow one to eliminate secondary structure as a variable. However,
using a homopolymer would not work for our purposes, because the size of
the polymer depends on ionic conditions. Isolating the role of
electrostatics requires an assembly template that not only has no
specific interactions with the coat protein but also has a fixed size
comparable to that of the native RNA.

\begin{figure*}
  \includegraphics{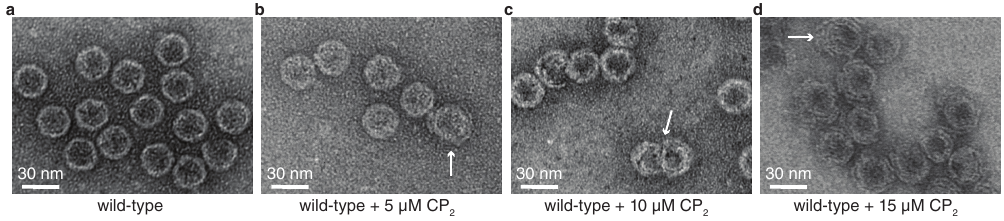}
  \caption{\label{fig:tem-layers} \textbf{Additional coat-protein layers
      can assemble on wild-type MS2 particles.} Unlike VLP assembly, in
    which capsids form around the viral RNA, here the wild-type MS2
    particle serves as the assembly template. Representative
    transmission electron microscopy (TEM) images show (a) wild-type MS2
    particles and (b--d) wild-type particles incubated with coat-protein
    dimers (CP$_2$) at (b) \SI{5}{\micro\molar}, (c)
    \SI{10}{\micro\molar}, and (d) \SI{15}{\micro\molar} concentration,
    all with negative staining. Arrow in (b) points to a particle with
    an incomplete secondary coat-protein layer. Arrow in (c) points to a
    doublet with a secondary coat-protein layer around both core
    particles. Arrow in (d) points to a particle with more than one
    additional coat-protein layer.}
\end{figure*}

Here we use the wild-type MS2 capsid itself as the template for
\textit{in vitro} assembly of MS2 coat proteins. Wild-type MS2 has
icosahedral symmetry with a triangulation number of $T = 3$ and consists
of 178 coat proteins (89 dimers; the maturation protein replaces one
coat-protein dimer in the icoahedral structure) that encapsidate its
native RNA,\cite{valegaard1990three, golmohammadi1993refined,
  koning2016asymmetric, dai2017insitu} along with a maturation protein
that binds to the receptor on the host, \textit{E.\@ coli}, and allows
the genome to enter it.\cite{lodish_mutants_1965,
  roberts_reconstitution_1967, sugiyama1967ribonucleoprotein,
  kozak_fate_1971, meng_structural_2019} Unlike homopolymers or RNA, the
structure of wild-type particles does not vary strongly with the ionic
strength of the solution. Furthermore, the capsid prevents additional
coat protein from interacting specifically with the RNA. However, coat
protein outside the capsid can still interact electrostatically with the
capsid and even the encapsidated RNA, if its charge is incompletely
screened by the capsid. We therefore expect the interactions between
coat protein in solution and the wild-type particles to be dominated by
electrostatics.

We show that additional layers of coat protein can assemble around
wild-type MS2 particles. We use a combination of transmission electron
microscopy (TEM), dynamic light scattering (DLS), gel electrophoresis,
zeta-potential measurements, and infection assays to assess the
morphology of these particles and understand how they form. Our results
are consistent with a nucleation-and-growth
model\cite{perlmutter2014pathways} for the assembly of additional
shells, in which electrostatic attraction provides the driving force for
assembly, while the nucleation barrier is controlled by the coat-protein
concentration.

\section{Results}

Our experimental approach is described in detail in \textit{Materials
  and Methods}. In brief, we combine \SI{50}{\nano\molar} wild-type MS2
particles with MS2 coat protein at concentrations from
\SIrange{1.25}{20}{\micro\molar}. Because coat proteins of MS2 and
related bacteriophages are thought to exist predominantly as dimers
under \textit{in vitro} assembly
conditions,\cite{beckett_ribonucleoprotein_1988, ni1995crystal} we
express all concentrations in terms of coat-protein dimers (CP$_2$). To
minimize specific interactions during assembly, we add RNase to digest
excess RNA released from broken capsids in the MS2 stock before the
start of the assembly experiments. After adding coat protein, we
incubate samples for \SI{10}{\minute}, a timescale based on previous
experiments\cite{garmann2019measurements} showing that assembly can take
on the order of \SI{1}{\minute} at these protein concentrations. We then
characterize the resulting assemblies using TEM to visualize particle
morphology, gel electrophoresis and DLS to infer size distributions,
zeta potential measurements to assess electrostatic interactions, and
plaque assays and fluorescence-based confocal microscopy to determine
the infectivity of the assembly products.

At coat-protein dimer concentrations above \SI{5}{\micro\molar}, we
observe the formation of additional, conformal layers around the
wild-type MS2 core (Fig.~\ref{fig:tem-layers} and
Fig.~\ref{fig:multiTEM}). We see no
secondary layers below this concentration. We note that a
\SI{5}{\micro\molar} concentration corresponds to a 100:1 ratio of
coat-protein dimers to wild-type particles (which are present at
\SI{50}{\nano\molar})---in principle, enough protein to make a second
$T=3$ capsid around each wild-type particle. However, at
\SI{5}{\micro\molar} concentration, only a small fraction of particles
have a secondary layer, and these layers do not fully surround the MS2
core particle (Fig.~\ref{fig:tem-layers}b). Nonetheless, this
observation shows that intact, wild-type MS2 particles can act as a
template for assembly.

The completeness of the secondary layers increases with increasing
protein concentration. At \SI{10}{\micro\molar}, a larger fraction of
particles are fully surrounded by an additional layer
(Fig.~\ref{fig:tem-layers}c). At \SI{15}{\micro\molar} and above, we see
that most particles are fully surrounded, with many having not only a
secondary protein layer, but also tertiary and higher-order layers
(Fig.~\ref{fig:tem-layers}d).

\begin{figure}
\includegraphics{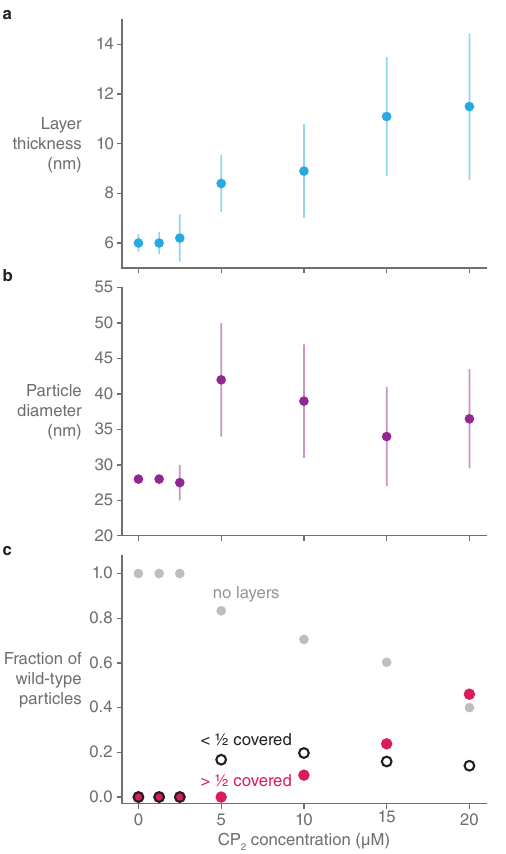}
\caption{\label{fig:tem-quantification} \textbf{Thickness and
    completeness of additional coat-protein layers increase with
    increasing coat-protein concentration.} Image analysis of multiple
  particles ($N=20$) at each coat-protein dimer concentration yields
  measurements of (a) average thickness of the capsid and additional
  layers; (b) average diameter; and (c) fraction of particles with no
  additional coat-protein layers (solid gray circles), additional layers
  that cover less than one half of the particle (open black circles),
  and additional layers that cover more than one half of the particle
  (solid red circles). Error bars for both layer thickness and particle
  diameter represent one standard deviation about the mean.}
\end{figure}

To quantitatively characterize the additional coat-protein layers that
form, we measure both the total thickness of the layer(s) and the
overall diameter of assembled particles directly from TEM images. We see
a sharp increase in both thickness (Fig.~\ref{fig:tem-quantification}a)
and overall diameter (Fig.~\ref{fig:tem-quantification}b) at
\SI{5}{\micro\molar} concentration, consistent with our observation that
secondary layers form only at \SI{5}{\micro\molar} concentration and
above. While the thickness increases with concentration above
\SI{5}{\micro\molar}---evidencing the growth of additional layers---our
measurements of the overall diameter of the particles do not reveal any
obvious trends, given the uncertainty in the measurement. We also count
the fraction of core particles with no layers, with additional layers
that cover less than one half of the particle, and with additional
layers that cover more than one half of the particle (see
Fig.~\ref{fig:classification} for examples and
Table~\ref{tab:concentration_tem} for data). These results show that the
completeness of additional coat-protein layers increases with
coat-protein concentration, though there are still some particles
without any secondary layers even at \SI{20}{\micro\molar}
(Fig.~\ref{fig:tem-quantification}c).

The TEM images also show some neighboring MS2 particles enclosed within
a common outer shell, producing doublets and other larger structures
(Fig.~\ref{fig:tem-layers}c). We observe these structures more
frequently in the TEM images as protein concentration increases.
However, sample preparation for electron microscopy involves dehydration
and staining that could influence particle morphology. We therefore
corroborate these results with measurements from gel electrophoresis and
DLS, which characterize the assembly products in solution.

\begin{figure}
  \centering
  \includegraphics{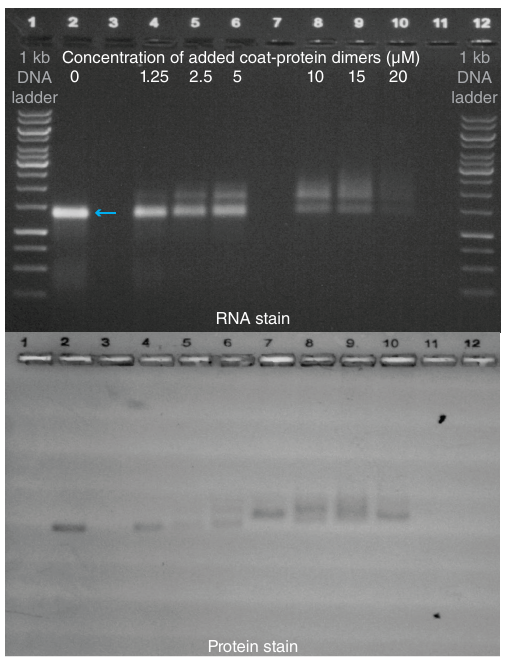}
  \caption{\label{fig:gel} \textbf{Gel electrophoresis shows that
      assembly products grow larger with increasing coat-protein
      concentration.} Photographs of agarose gel containing assembly
    products from mixing \SI{50}{\nano\molar} wild-type MS2 particles
    with \SIrange{0}{20}{\micro\molar} coat-protein dimers
    (concentration for each lane is shown in image overlay; lane 7
    contains a sample with an error in composition and can be ignored).
    Top image shows gel stained with ethidium bromide for labeling
    nucleic acid, and bottom image shows gel stained with Coomassie Blue
    R-250 for labeling protein. The blue arrow points to the primary
    band, which corresponds to the wild-type capsid size at
    \SI{0}{\micro\molar} concentration, but gradually shifts to lower
    mobilities at higher concentrations. This shift is evidenced by the
    position of the band relative to the tenth band (from top) of the DNA
    ladder. A secondary band can be seen at all non-zero concentrations.
    This band, which becomes more diffuse with increasing protein
    concentration, likely represents doublets of particles or larger
    agglomerates that are bound together by the added coat protein.  }
\end{figure}

The results from gel electrophoresis are consistent with the TEM
observations. We see a primary band that appears under both RNA- and
protein-staining, identifying it as containing encapsidated structures
(Fig.~\ref{fig:gel}). As protein concentration increases, the band
decreases in intensity in the RNA-stained image and increases in the
protein-stained image. The mobility of the band also decreases as
protein concentration increases, as indicated by the slight slowing
relative to wild-type capsids and to the DNA ladder. This decrease is
consistent with an increase in average size due to additional layers of
protein, but the degree to which the band spreads is less than what we
would expect from the TEM images, which show that many particles have no
coverage. The increase in intensity in the protein-stained image gives
stronger evidence that the particles have secondary layers.

A second band is visible at low protein concentrations and becomes
increasingly prominent until it starts to become diffuse at the highest
concentrations. Although this second band also appears under both RNA
and protein staining, its clear separation from the primary band
suggests that it does not contain individual particles with secondary
layers, since such particles are unlikely to be so well differentiated
from wild-type capsids. Instead, the second band likely contains
doublets of particles surrounded by a single secondary layer, as seen in
the TEM images. The increasing diffuseness of the second band at higher
concentrations is consistent with the formation of even larger
agglomerates of MS2 particles, which are also seen in the TEM images.

\begin{figure}
  \centering\includegraphics{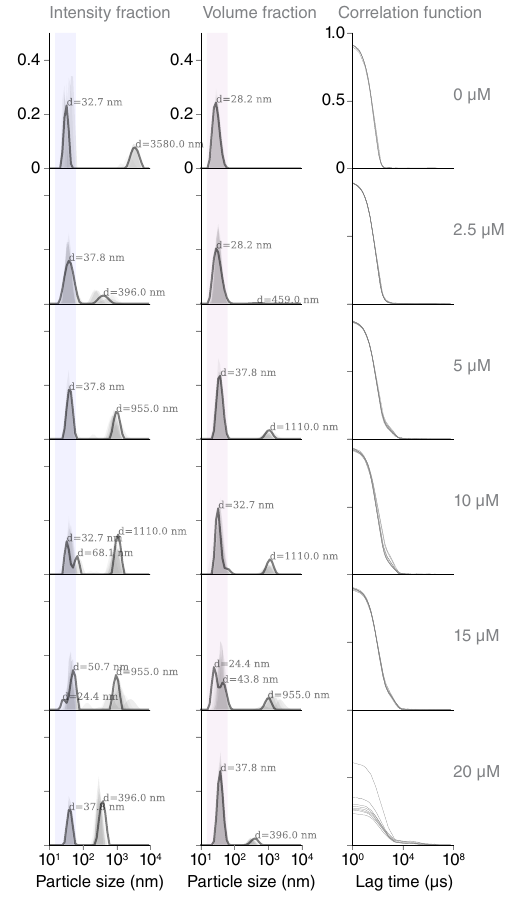}
  \caption{\label{fig:dls} \textbf{Dynamic light scattering shows an
      increase in particle size with increasing protein concentration.}
    First two columns show plots of size distributions inferred from DLS
    measurements of the assembly products of \SI{50}{\nano\molar}
    wild-type MS2 particles mixed with \SIrange{0}{20}{\micro\molar}
    coat-protein dimers (concentrations are shown at right). The first
    column shows the distribution on an intensity basis, and the second
    shows the distribution on a volume basis. Light gray peaks in the
    distributions show the results from individual measurements, while
    the dark gray peaks show the results inferred from the average
    autocorrelation function. Blue and purple shaded regions show the
    range of diameters consistent with wild-type particles. The diameter
    $d$ corresponding to each peak is noted on each distribution. The
    autocorrelation function for each measurement is shown in the plots
    in the third column.}
\end{figure}

The sizes that we infer from DLS measurements are less well-resolved
than the gel electrophoresis measurements but span a broader range. The
volume-based size distribution of wild-type MS2 shows a peak at the
expected size of native capsids (about \SI{28}{\nano\meter}), as shown
in Fig.~\ref{fig:dls}. Samples prepared at \SI{2.5}{\micro\molar}
concentration show a similar size distribution, with a small peak
appearing at much larger sizes. As the protein concentration increases
to \SI{5}{\micro\molar} and above, the primary peak appears to shift to
larger diameters or to bifurcate. Meanwhile, the amplitude of the peak
at larger sizes increases in the intensity-based size distribution but
remains approximately constant in the volume-based distribution,
indicating that the majority of particles remain comparable in size to
wild-type MS2 (see purple shaded region of volume distributions in
Fig.~\ref{fig:dls}).

The changes to the primary peak at \SI{5}{\micro\molar} protein
concentration and above provide qualitative evidence for changes to the
structure of the wild-type particles. We say ``qualitative'' because we
do not expect DLS to be able to resolve the broadening in the size
distribution associated with the formation of a secondary protein layer
on individual (unaggregated) wild-type particles. The inability of DLS
to resolve finely spaced modes in the size distribution arises because
inverting the correlation function is an ill-conditioned inverse
problem. Solving this problem robustly requires assuming the
distribution to be smooth,\cite{provencher1982constrained,
  provencher1982contin, scales1993data} which results in an artificial
broadening of even highly monodisperse distributions. With this caveat
in mind, we note that the bifurcation of the primary peak seen in the
intensity-based data at \SI{10}{\micro\molar} and \SI{15}{\micro\molar}
concentration is consistent with the gel data showing two bands at
elevated concentration. The larger of the two bifurcated peaks in the
DLS data, which occurs at a size of about \SI{50}{\nano\meter} or
\SI{60}{\nano\meter}, is consistent with doublet formation. The peak at
large sizes may reflect the presence of much larger aggregates or dust,
which would not be easily detected in the gel experiments.

\begin{figure}
\centering\includegraphics{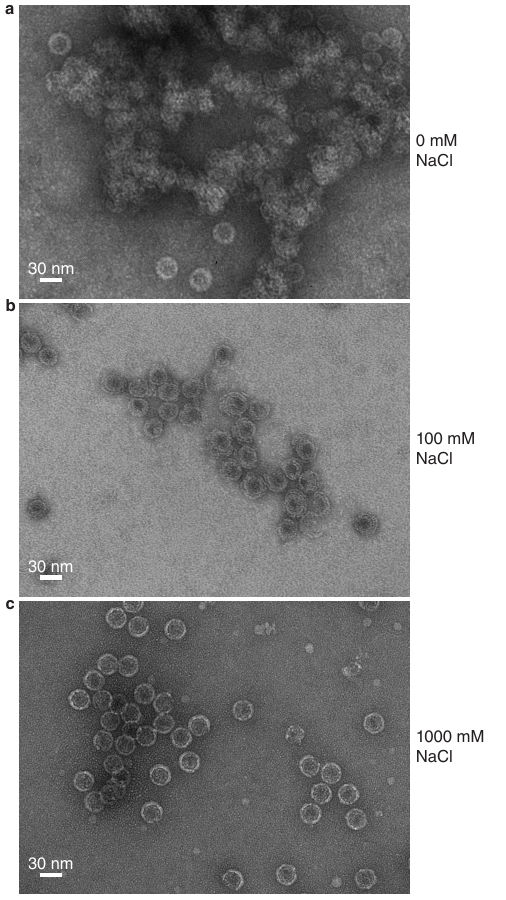}
\caption{\label{fig:salt} \textbf{Additional coat-protein layers
    preferentially form at moderate salt concentration.} Transmission
  electron microscopy images of \SI{50}{\nano\molar} wild-type MS2 mixed
  with \SI{15}{\micro\molar} coat-protein dimers in buffer containing
  (a) \SI{0}{\milli\molar}, (b) \SI{100}{\milli\molar} (same image as in
  Fig.~\ref{fig:tem-layers}d), and (c) \SI{1000}{\milli\molar} NaCl.
  Only at \SI{100}{\milli\molar} NaCl do we see clear evidence for
  particles with additional coat-protein layers.}
\end{figure}

To understand the mechanism driving the formation of the additional
coat-protein layers, we examine how the assembly products change as the
salt concentration changes. In these experiments we combine
\SI{50}{\nano\molar} wild-type MS2 with \SI{15}{\micro\molar}
coat-protein dimers at NaCl concentrations from
\SIrange{0}{1000}{\milli\molar}. We expect salt concentration to affect
both the protein-protein interactions as well as the RNA-protein
interactions.

We find clear evidence for secondary layer formation only at salt
concentrations around \SI{100}{\milli\molar}. At \SI{0}{\milli\molar}
NaCl, we do not observe the formation of additional capsid shells;
instead, we observe aggregates of the virus particles
(Fig.~\ref{fig:salt}a). At \SI{100}{\milli\molar} NaCl, we see multiple
additional coat-protein layers (Fig.~\ref{fig:salt}b), as shown
previously in Fig.~\ref{fig:tem-layers}d. At \SI{1000}{\milli\molar}
NaCl, the TEM images show particles with inhomogeneous protein shells
(Fig.~\ref{fig:salt}c). We see similar inhomogeneity at
\SI{300}{\milli\molar} NaCl. Notably, at \SI{600}{\milli\molar} we
directly observe protein aggregation happening on the wild-type cores.
It is therefore possible that aggregation of coat-protein onto the
wild-type cores is responsible for the inhomogeneities observed at salt
concentrations above \SI{100}{\milli\molar}. It is also possible that
crystallization of the NaCl during drying for TEM is responsible for the
inhomogeneities. Electron micrographs for samples at
\qtylist{50;300;600}{\milli\molar} are shown in
Fig.~\ref{fig:multinacl}.

\begin{table}
  \caption{\label{tab:zeta} Zeta potentials for assembly components and
    products, as measured with dynamic light scattering. Uncertainties
    represent the standard deviation across multiple runs.}
  \begin{ruledtabular}
    \begin{tabular}{l r S[table-format=2.2(2), separate-uncertainty] }
      Sample                             & NaCl (\unit{\milli\molar}) & {Zeta potential (\unit{\milli\volt})} \\
      \colrule
      coat-protein dimers (CP$_2$)       & 100                        & 6.97 +- 0.23   \\
      wild-type MS2                      & 100                        & -16.88 +- 1.28 \\
      wild-type MS2                      & 1000                       &  2.78 +- 3.85   \\
      MS2 + \SI{15}{\micro\molar} CP$_2$ & 0                          & -17.4 +- 0.81  \\
      MS2 + \SI{15}{\micro\molar} CP$_2$ & 100                        & -9.44 +- 0.12  \\
      MS2 + \SI{15}{\micro\molar} CP$_2$ & 1000                       & -7.50 +- 2.7   \\
  \end{tabular}
  \end{ruledtabular}
\end{table}

For completeness, we measure the apparent zeta potential of coat-protein
dimers, wild-type particles, and products of assembling these components
at various salt concentrations (Table~\ref{tab:zeta}). We note that due
to the permeability of the capsid, there are fundamental difficulties in
interpreting the electrophoretic mobilities of MS2 particles in terms of
zeta potentials or surface charge.\cite{langlet_impact_2008,
  dika2011impact} Therefore the measurements we present are best
interpreted as indicators of how the particles respond to an electric
field at various salt concentrations. We find that at
\SI{100}{\milli\molar} NaCl, coat-protein dimers (\SI{15}{\micro\molar})
have a positive measured zeta potential, and wild-type MS2
(\SI{50}{\nano\molar}) has a negative potential. The apparent negative
zeta potential for the wild-type particle has been reported
previously\cite{vodolazkaya_estimation_2022} and may reflect the
influence of the RNA on the electrokinetic
properties.\cite{dika2011impact} We also find that the apparent zeta
potential of the assembly products at \SI{100}{\milli\molar} NaCl is
smaller (in magnitude) than that of wild-type capsids but still
negative. At \SI{1000}{\milli\molar} NaCl, the wild-type capsids have a
near-zero apparent zeta potential, while the assembly products have a
negative potential comparable to that at \SI{100}{\milli\molar}. These
results show that there are strong variations of the electrokinetic
properties with salt concentration. While it is not clear whether these
variations are related to changes in surface charge, the complex
electrokinetic behavior suggests that it is not unreasonable to expect
the assembly morphologies to vary markedly with salt concentration.

\begin{figure}
\centering\includegraphics{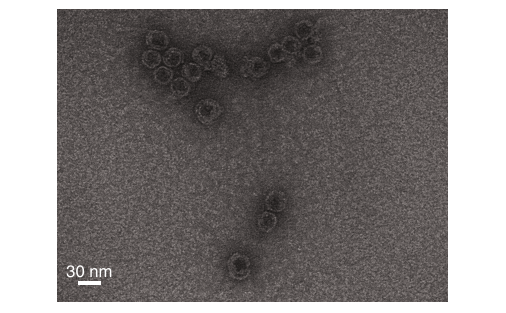}
\caption{\label{fig:vlp} \textbf{Additional layers assemble on VLPs.}
  Transmission electron microscopy image of assembly products from
  mixing \SI{50}{\nano\molar} MS2 with \SI{15}{\micro\molar}
  coat-protein dimers in \SI{100}{\milli\molar} NaCl.}
\end{figure}

The variations in layering of the assembly products and in their
apparent zeta potential with salt concentration strongly suggest that
electrostatic attraction is responsible for the assembly of secondary
layers, but it is also possible that the maturation protein, located on
the surface of wild-type MS2 particles, might play a role in nucleating
a secondary layer. To investigate this hypothesis, we assemble coat
protein around MS2 VLPs that do not have the maturation protein. We make
MS2 VLPs by mixing \SI{50}{\nano\molar} MS2 RNA and \SI{5}{\micro\molar}
coat-protein dimers. We then add RNase to ensure excess RNA is not
present. Finally, we combine the VLP sample with \SI{15}{\micro\molar}
coat-protein dimers at \SI{100}{\milli\molar} NaCl.

We find that added coat protein can form secondary layers around VLPs,
just as it can around wild-type MS2 particles (Fig.~\ref{fig:vlp}). The
diameter of the assemblies formed around the VLPs, as measured by
analysis of the TEM images, is \SI{32.7 +- 4.9}{\nano\meter}, which does
not differ significantly from that of the assemblies formed around
wild-type MS2 particles, \SI{34.4 +- 5.4}{\nano\meter}.

\begin{figure}
\centering\includegraphics{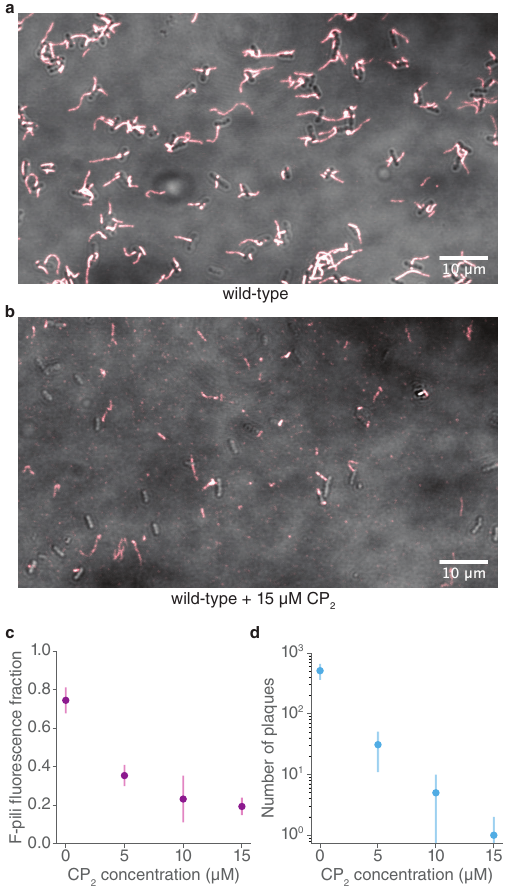}
\caption{\label{fig:infection} \textbf{Additional coat-protein layers
    block the maturation protein on the wild-type core particle.} (a)
  Confocal micrograph of AF647-labeled wild-type MS2 mixed with
  \textit{E.\@ coli}. The bacteria are visible in the bright-field
  channel, while the F-pili are made visible by the fluorescent MS2
  particles that have attached to them. (b) As in (a), but here the
  wild-type particles are first incubated with \SI{15}{\micro\molar}
  coat-protein dimers and then, following assembly, added to the
  bacterial suspension. (c) Plot of fluorescence fraction, defined as
  the fraction of AF647 fluorescence associated with F-pili, as a
  function of coat-protein dimer concentration. (d) Plot of number of
  plaques (from plaque assays) as a function of coat-protein dimer
  concentration. Photographs of plaque assays and additional confocal
  micrographs are shown in Fig.~\ref{fig:multiinfectivity}. The
  reduction in fluorescence and number of plaques both indicate reduced
  binding between the assembled particles and the F-pili, which suggests
  that some of the secondary layers block the maturation protein.}
\end{figure}

These results show that the maturation protein is not essential for
assembly of secondary layers but do not rule out the possibility that it
may aid in their formation when it is present. To further understand
whether the maturation protein plays a role in the assembly of secondary
layers, we observe how particles bind to and infect \textit{E.\@ coli}.
Because the maturation protein binds to the bacterial F-pilus in the
initial stages of infection, we expect particles whose maturation
protein is buried beneath an additional coat-protein layer to show
reduced binding and reduced infectivity.

To detect binding, we label wild-type MS2 with Alexa Fluor 647 (AF647),
then assemble in the presence of \qtylist{0;5;10;15}{\micro\molar}
coat-protein dimers. We find that additional coat-protein layers are
able to form after incubating the labeled particles for \SI{10}{\minute}
(Fig.~\ref{fig:multiTEMlabelledwt}). We then mix the assembly products
with \textit{E.\@ coli} cells and image under a confocal fluorescence
microscope.

Confocal images show that F-pili are much less well-decorated by
particles assembled at \SI{15}{\micro\molar} coat-protein than by
wild-type particles (Fig.~\ref{fig:infection}a--b and
Fig.~\ref{fig:multiinfectivity}). To
quantify the binding to the host cells, we measure the fluorescence
fraction, defined as the fraction of total AF647 fluorescence associated
with the F-pili. We find that the fluorescence fraction decreases with
increasing protein concentration (Fig.~\ref{fig:infection}c),
indicating that secondary layers occlude the maturation protein more
frequently at higher protein concentrations. Because some particles can
still attach to the F-pili even at high protein concentration, we infer
that the secondary layers are randomly situated on the capsid surface
relative to the maturation protein; that is, there is no evidence that
the secondary layers preferentially assemble at the maturation protein.

Because fluorescent labeling of the capsid might affect the assembly of
additional layers and therefore the binding to F-pili, we also perform
plaque assays with unlabeled MS2. The assays show that infectivity
decreases with the protein concentration used to assemble the particles
(Fig.~\ref{fig:infection}d). These findings are consistent with the
confocal measurements.

\section{Discussion}

Our experiments show that MS2 coat protein can assemble on wild-type MS2
particles. As the added protein concentration increases, the
completeness of the secondary layers also increases, with more complex
multilayer assemblies seen at the highest protein concentrations.
Furthermore, these secondary coat-protein layers form only at salt
concentrations around \SI{100}{\milli\molar} NaCl. Taken together, these
results indicate that the driving force for assembly of the secondary
layers is electrostatic attraction between positively charged coat
proteins and the negatively charged wild-type particle.

The mechanism for the formation of additional layers appears to differ
from that of CCMV, the coat protein of which can form capsids with
multiple layers near the isoelectric point of the capsid
exterior\cite{lavelle2009phase}. Our experiments, however, are conducted
at neutral pH, well above the isoelectric point of MS2
(pH~3.9).\cite{michen2010isoelectric} Moreover, in CCMV, additional
layers assemble in equilibrium\cite{lavelle2009phase,
  prinsen_multishell_2010} in the absence of RNA, whereas in our system,
additional layers form only when coat-protein is added to pre-formed
particles (either wild-type MS2 or VLPs). We note that when we mix
\SI{15}{\micro\molar} coat-protein dimers in assembly buffer in the
absence of pre-formed particles, we do not observe any assembled
structures, showing that the initial particles are necessary to promote
the assembly of multilayered particles.

Combined with the observation of a concentration threshold for the
appearance of secondary layers, the necessity of the seed particles
strongly suggests that additional layers form through heterogeneous
nucleation and growth. Previous single-particle imaging
studies\cite{garmann2019measurements} and bulk
experiments\cite{williams2024effect} have shown that MS2 VLPs assemble
through a nucleation-and-growth pathway under much the same solution
conditions as we use here. However, in our experiments, the pre-formed
particles act as the substrate, rather than the RNA. The previous
experiments also showed that the nucleation rate depends strongly on
coat-protein concentration, leading to a concentration threshold below
which nucleation does not occur over the timescale of the experiment. In
those experiments, the concentration threshold was about
\SI{2}{\micro\molar}. In our experiments, we observe no secondary layers
below \SI{5}{\micro\molar}. This higher threshold points to a larger
nucleation barrier for assembly around pre-formed particles than around
RNA---which is sensible, considering that the RNA is encapsidated in our
experiments. The consequent elimination of sequence-specific
interactions---for example, with the TR loop---should decrease the
energetic driving force for coat protein to bind to the template.

By contrast, increasing the coat-protein concentration should reduce the
entropic penalty for protein to bind to the particle, thereby reducing
the nucleation barrier. We therefore surmise that the particles with
multiple layers we observe at high protein concentration
(Fig.~\ref{fig:tem-layers}d) form when multiple nuclei grow on
the same particle.

A major difference between the coat-protein layers we observe here and
those assembled around RNA is that our secondary layers are incomplete.
Even at high protein concentration (\SI{15}{\micro\molar} dimers), TEM
images show many particles with no layers or partial layers
(Fig.~\ref{fig:tem-quantification}c), and confocal measurements show
that many multilayer particles can still bind to F-pili
(Fig.~\ref{fig:infection}c), indicating that the maturation protein
remains unoccluded. If the assembly of secondary layers is nucleated, as
we claim, it may seem strange that their growth does not proceed to
completion.

However, the growth of secondary layers incurs two types of
elastic-energy penalties. First, compared to RNA, a capsid is a more
rigid substrate for assembly, which should increase the energetic cost
to encapsulate it.\cite{li2017effect} Second, the curvature of the
secondary layer differs from that of the first layer (the capsid). When
multiple layers co-assemble in equilibrium, as is the case in
CCMV\cite{lavelle2009phase}, this elastic strain is offset by
electrostatic attractions\cite{prinsen_multishell_2010}. In our
experiments, however, the secondary layers assemble on a template with
fixed curvature. In other soft materials, including
vesicles\cite{schneider_shapes_2005} and colloid-covered
droplets\cite{meng_elastic_2014}, a mismatch between the curvature of
the template and the preferred curvature of the assembling species can
lead to an elastic instability that arrests growth. A similar
instability might be at work in our system as well.

This argument does not explain why doublets and larger aggregates of
particles form with increasing frequency as protein concentration
increases. We would expect a secondary layer that covers more than one
particle to have an even smaller curvature, and hence larger elastic
energy cost, than a layer that covers one. However, previous simulations
of the assembly of proteins around RNA at large excess protein
concentration showed that a partial capsid growing on one RNA molecule
can connect to a partial capsid on another RNA molecule, bridging
them.\cite{williams2024effect} An analogous process might occur when the
assembly template is a particle rather than the RNA, and when secondary
layers are assembling rather than capsids. Further experiments that
image structure at higher resolution may be able to shed light on how
these larger structures form, as well as how assembly goes awry at low
and high salt concentrations.

\section{Conclusion}

Under \textit{in vitro} solution conditions very similar to those under
which MS2 coat-protein assembles around MS2 RNA, MS2 coat protein can
assemble into secondary layers around wild-type MS2 particles. In both
cases, the assembly appears to be nucleated, but in the case of the
secondary layers, the driving force is reduced because specific
interactions are absent and the layers are strained. Consequently, the
secondary layers form at a larger threshold protein concentration, and
the elastic strain may also prevent them from growing to completion.
Nonetheless, the electrostatic interaction between the coat proteins and
the capsid is sufficient to drive formation of additional coat-protein
layers. These results highlight the importance of electrostatics in the
assembly of protein shells, even in systems such as MS2 that are known
for their strong, specific RNA-protein interactions.

From a practical perspective, our results might help in the design of
VLPs. For some applications, such as drug delivery,~\cite{kovacs07,
  galaway13, hartman2018quantitative} multiple layers might help to
increase the stability of the particle \textit{in vivo}. From a
fundamental perspective, layered MS2 particles might serve as a model
experimental system for exploring scaffolded assembly.\cite{li2018large}

\section*{Acknowledgments}
We thank Amy Barker and Peter Stockley at the University of Leeds for
initial stocks of MS2 and \textit{E.\@ coli} cells. This research was
partially supported by a National Science Foundation (NSF) Graduate
Research Fellowship under grant number DGE-1745303 and by NSF through
the Harvard University Materials Research Science and Engineering Center
under NSF grant number DMR-2011754, by the National Institute of General
Medical Sciences of the National Institutes of Health under grants
K99/R00GM127751 and R35GM157105, by the NSF-Simons Center for
Mathematical and Statistical Analysis of Biology at Harvard University
under NSF grant number 1764269, and by the Harvard Quantitative Biology
Initiative. This work was performed in part at the Harvard University
Center for Nanoscale Systems (CNS), a member of the National
Nanotechnology Coordinated Infrastructure Network (NNCI), which is
supported by the National Science Foundation under NSF grant number
ECCS-2025158. The work was also performed in part at the Harvard
University Bauer Core Facility.

\section*{Author declarations}
\subsection*{Conflicts of interest }
There are no conflicts to declare.

\subsection*{Author contributions}
All authors conceived the research goals and aims. LAW, AP, and KK
curated data. LAW and AP analyzed the results, with assistance from VNM.
RFG and VNM obtained funding and supervised the project. LAW, AP, and KK
performed the experiments. All authors developed methods. LAW and VNM
prepared visualizations. LAW and VNM prepared the original draft, and
all authors reviewed and edited the submitted work.

\section*{Data availability}
Upon acceptance, experimental data will be made freely available at the
Harvard Dataverse.

\section*{Materials and Methods}\label{sec:methods}

\subsection*{Assembly experiments}
Protocols for virus growth, cultivation, and storage; coat-protein
purification and storage; and RNA purification and storage are given in
Ref.~\citenum{williams2024effect}. The virus protocols follow those of
Strauss and Sinsheimer,\cite{strauss1963purification} and the
coat-protein protocols follow those of Sugiyama, Herbert, and
Hartman.\cite{sugiyama1967ribonucleoprotein}

For the assembly experiments around wild-type particles, we mix purified
wild-type MS2 at a concentration of \SI{50}{\nano\molar} with purified
MS2 coat-protein dimers at concentrations from
\SIrange{1.25}{20}{\micro\molar}. For the assembly experiments around
VLPs, we first make VLPs by mixing MS2 RNA at \SI{50}{\nano\molar}
concentration with purified coat-protein dimers at \SI{5}{\micro\molar}.
We add \SI{10}{\nano\gram} RNase A and wait \SI{30}{\minute} to allow
any RNA that is not encapsidated to be digested. We then combine the VLP
sample with \SI{15}{\micro\molar} coat-protein dimers. All assembly
experiments are done in assembly buffer (\SI{42}{\milli\molar} Tris, pH
7.5; \SI{3}{\milli\molar} acetic acid, \SI{1}{\milli\molar} EDTA, with
NaCl concentrations as specified in the text) at room temperature
(\SI{21}{\celsius}) for \SI{10}{\minute}.

\subsection*{Transmission electron microscopy and quantitative analysis}
For TEM imaging, we first negatively stain samples with 2\% aqueous
uranyl acetate on 200 mesh carbon-coated copper TEM grids (Polyscience,
TEM-FCF200CU). We then image the negatively stained samples using a
Hitachi 7800 TEM (Center for Nanoscale Systems in the Science and
Engineering Complex at Harvard University) at \qtylist{20; 50;
  100}{\kilo\volt}.

We analyze TEM images using Fiji (ImageJ). For each micrograph, we
calculate the pixel-to-nanometer calibration factor using the image
scale bar. We use the Fiji line measurement tool to measure the diameter
and coat-protein layer thickness for $N=20$ randomly selected particles
at each experimental condition. We then calculate the mean and
sample standard deviation for each condition.

\subsection*{Gel electrophoresis and analysis}
For gel electrophoresis we prepare 1\% agarose gels in a mixture of 5
parts TNE buffer (\SI{50}{\milli\molar} Tris, \SI{100}{\milli\molar}
NaCl, \SI{10}{\milli\molar} EDTA, pH 7.5) and 1 part
\SI{20}{\milli\molar} acetic acid buffer. We load the gels with a
mixture of \SI{15}{\micro\liter} sample and \SI{4}{\micro\liter}
glycerol. For RNA staining we use ethidium bromide, and for protein
staining we use Coomassie Blue R-250. After loading, we run the gels at
room temperature at \SI{100}{\volt} for \SI{40}{\minute}. We image the
gels with a Biosystems UV Imager (Azure AZ1280).

\subsection*{Dynamic light scattering and zeta potential measurements}
We use a Malvern ZetaSizer Nano ZS by Malvern Panalytical for both
dynamic light scattering and zeta potential measurements. To determine
size distributions from the correlation functions, we use the
regularized inversion method provided with the instrument's
software.\cite{zeta07} We prepare samples for zeta-potential
measurements at the solution conditions discussed in the main text. The
concentration of the coat-protein dimer sample was
\SI{15}{\micro\molar}, and the concentration of the wild-type MS2
samples was \SI{50}{\nano\molar}.

\subsection*{Fluorescent labeling and confocal microscopy}
We fluorescently label wild-type MS2 with Alexa Fluor 647 succinimidyl
ester (Thermo Fisher Scientific A37573). We incubate the virus for
\SI{1.5}{\hour} in \SI{0.1}{\molar} sodium bicarbonate and
\SI{1}{\milli\gram\per\milli\liter} dye. We wash the dyed particles with
TNE buffer (\SI{50}{\milli\molar} Tris, pH 7.5; \SI{100}{\milli\molar}
NaCl, \SI{1}{\milli\molar} EDTA) using \SI{100}{\kilo\dalton} MWCO
centrifugal filters (Millipore Sigma, UFC100824) and store at
\SI{4}{\celsius}.

For confocal imaging, we place \SI{10}{\micro\liter} of \textit{E.\@
  coli} C3000 cell culture on a base-washed No.~1 coverslip (Globe
Scientific, 1419-10) and add \SI{2}{\micro\liter} labeled particles. To
immobilize the cells, we place a 2\% agarose pad made with PBS buffer on
top of the sample. We image these samples on a Leica TCS SP5 confocal
microscope with a 63$\times$ water-immersion objective (NA 1.2) and a
\SI{633}{\nano\meter} excitation wavelength. We also acquire bright-field
images of the same fields of view by imaging the light transmitted
through the sample. This technique allows us to visualize the location
and morphology of individual \textit{E.\@ coli} cells in the transmitted
channel while also detecting the fluorescently labeled MS2 particles.

To calculate the fluorescence fraction, we first identify individual
F-pili in each confocal image and mask them. We then calculate the total
fluorescence intensity within the mask, $I_{\text{pili}}$, and outside
the mask, $I_{\text{non-pili}}$, from the AF647 fluorescence channel. To
account for background fluorescence inside the mask, we estimate the
mean fluorescence intensity outside the mask and subtract the
corresponding background contribution from the measured pili intensity.
The background-corrected pili fluorescence is
\begin{equation}
I_{\text{pili,corr}} = I_{\text{pili}} - \left(\frac{N_{\text{pili}}}{N_{\text{non-pili}}}\right)
I_{\text{non-pili}},
\end{equation}
where $N_{\text{pili}}$ and $N_{\text{non-pili}}$ are the numbers of
pixels within and outside the pili mask. We then calculate
the fluorescence fraction as
\begin{equation}
  f = \frac{I_{\text{pili,corr}}}{I_{\text{pili}}+I_{\text{non-pili}}}.
\end{equation}
We calculate the fluorescence fraction independently for $N=3$ images
at each CP$_2$ concentration and report the mean and standard deviation
across the three images for each concentration.

\subsection*{Plaque assay}
We dilute each assembly product (and the wild-type particles) by a
factor of $10^6$, mix \SI{200}{\micro\liter} \textit{E.\@ coli} into
each of the diluted samples, add \SI{10}{\milli\liter} top agarose
(\SI{2}{\gram} Tryptone; \SI{0.2}{\gram} yeast extract; \SI{2.92}{\gram}
sodium chloride; \SI{0.9}{\gram} agarose; \SI{200}{\milli\liter} Milli-Q
water), and pour into a
\SI{90}{\milli\meter}$\times$\SI{17}{\milli\meter} cell culture plate
(VWR, 10062-878) containing \SI{15}{\milli\liter} bottom agar
(\SI{3}{\gram} Tryptone; \SI{0.3}{\gram} yeast extract; \SI{2.4}{\gram}
sodium chloride; \SI{4.5}{\gram} agar; \SI{300}{\milli\liter} Milli-Q
water). We let the plates cool for \SI{5}{\minute} and place them upside
down into an incubator at \SI{37}{\celsius} overnight. During this time,
the \textit{E.\@ coli} grows on top of the bottom agar. Particles that
infect the \textit{E.\@ coli} create a plaque---or hole---with each
plaque representing (in principle) an infection starting from a single
infectious particle.

\section*{References}

\bibliography{bibliography}


\clearpage
\appendix
\setcounter{figure}{0}
\setcounter{table}{0}
\renewcommand{\thefigure}{S\arabic{figure}}
\renewcommand{\thetable}{S\arabic{table}}
\renewcommand{\theHfigure}{appendixS-\arabic{figure}}
\renewcommand{\theHtable}{appendixS-\arabic{table}}

\begin{figure*}[t]
\centering\includegraphics[width=0.95\textwidth]{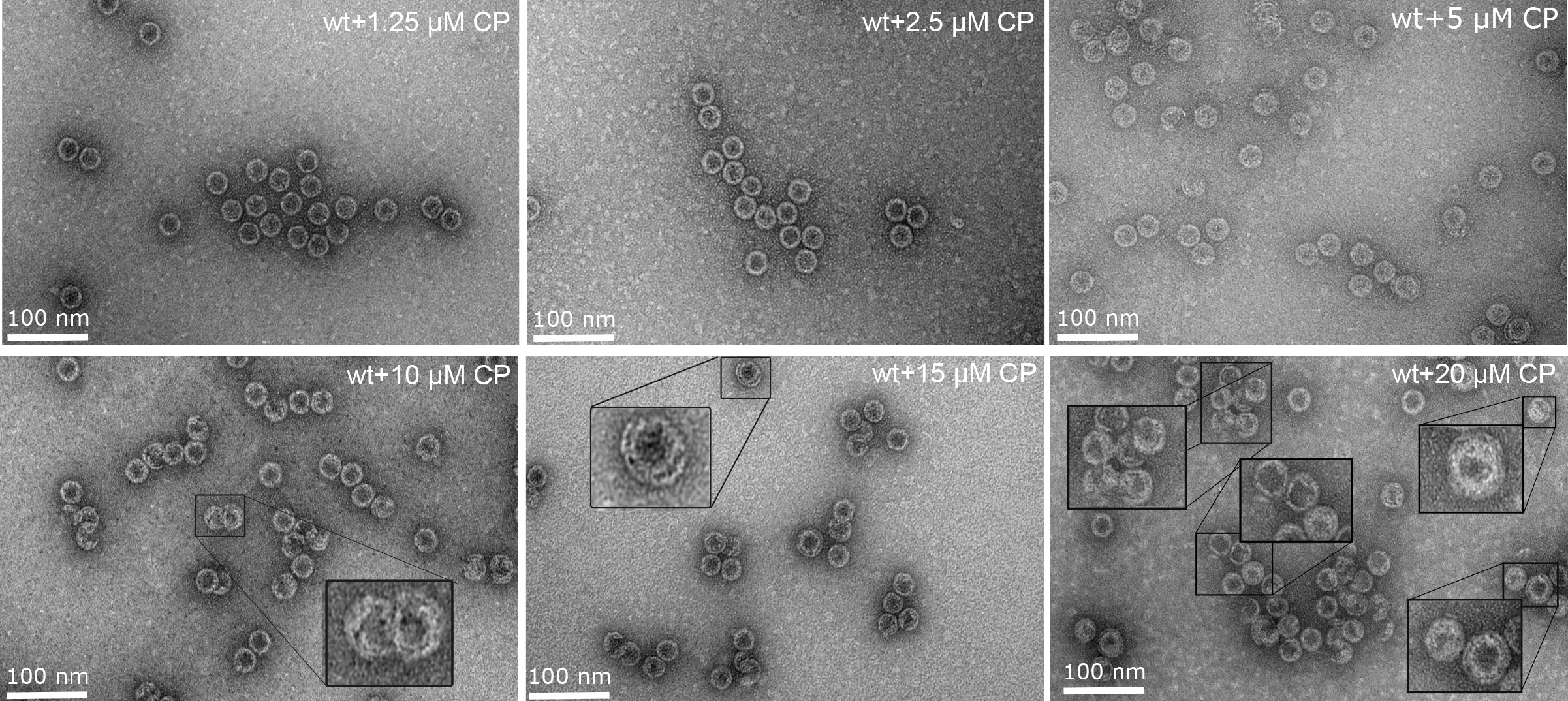}
\caption{\label{fig:multiTEM}Additional TEM images of negatively stained
  samples of wild-type MS2 particles mixed with varying coat-protein
  dimer concentrations. }
\end{figure*}

\begin{figure*}[t]
\centering\includegraphics[width=0.95\textwidth]{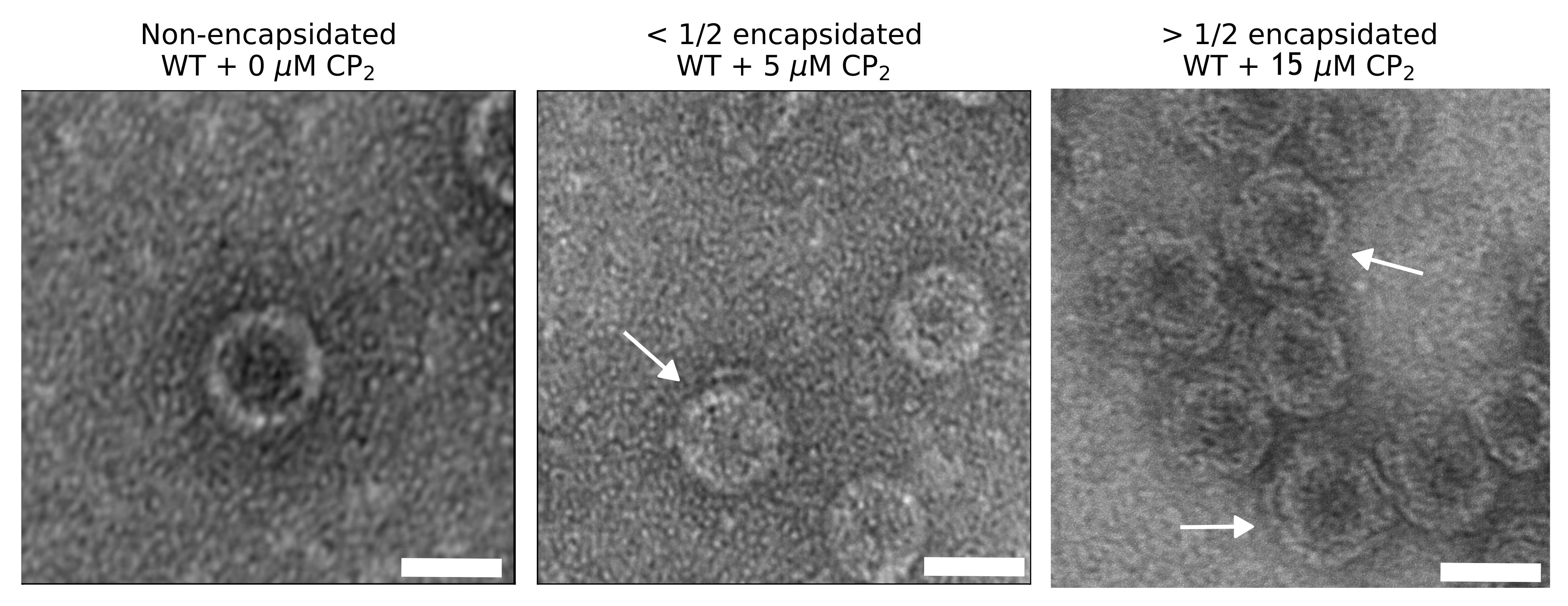}
\caption{\label{fig:classification} Representative TEM images
  illustrating classification of secondary coat-protein encapsidation
  around wild-type MS2. Representative particles were classified as
  non-encapsidated when no additional coat-protein shell was observed,
  $<1/2$ encapsidated when an additional coat-protein layer covered less
  than approximately half of the WT MS2 particle, and $>1/2$
  encapsidated when the additional layer extended around more than
  approximately half of the particle. Complete secondary shells and
  particles displaying multiple concentric coat-protein shells were
  classified as $>1/2$ encapsidated. Scale bars 30 nm.}
\end{figure*}

\begin{table*}
  \caption{\label{tab:concentration_tem} Layer thickness, particle
    diameter, and degree of coat-protein encapsidation for wild-type MS2
    incubated with increasing concentrations of MS2 coat-protein dimers
    (CP$_2$). Approximately $N=20$ particles were analyzed for each
    condition. Encapsidation is reported as the number of particles out
    of 20 randomly chosen particles classified as $<1/2$ encapsidated or
    $>1/2$ encapsidated.}

  \begin{ruledtabular}
    \begin{tabular}{
      rS[table-format=2.2(3), separate-uncertainty] S[table-format=2.3(3), separate-uncertainty] c c}
      {CP$_2$ concentration} & {Layer Thickness (\unit{\nano\meter})} & {Particle Diameter (\unit{\nano\meter})} & {$<1/2$ Encapsidated} & {$>1/2$ Encapsidated} \\

    \colrule
    0~$\mu$M     & 6.0 +- 0.01  & 28.3 +- 0.01 & $0/20$ & $0/20$ \\
    1.25~$\mu$M  & 6.0 +- 0.1  & 27.4 +- 0.6  & $0/20$ & $0/20$ \\
    2.5~$\mu$M   & 6.2 +- 1.0  & 27.5 +- 1.8 & $1/20$ & $0/20$ \\
    5~$\mu$M     & 8.5 +- 1.3 & 41.9 +- 8.0  & $3/20$ & $1/20$ \\
    10~$\mu$M    & 8.9 +- 1.9  & 38.9 +- 7.9  & $5/20$ & $3/20$ \\
    15~$\mu$M    & 11.1 +- 2.4 & 34.1 +- 5.4  & $4/20$ & $8/20$ \\
    20~$\mu$M    & 11.6 +- 2.9 & 36.2 +- 5.9  & $6/20$ & $4/20$ \\
    \end{tabular}
  \end{ruledtabular}
\end{table*}

\begin{figure*}[t]
  \centering\includegraphics[width=0.95\textwidth]{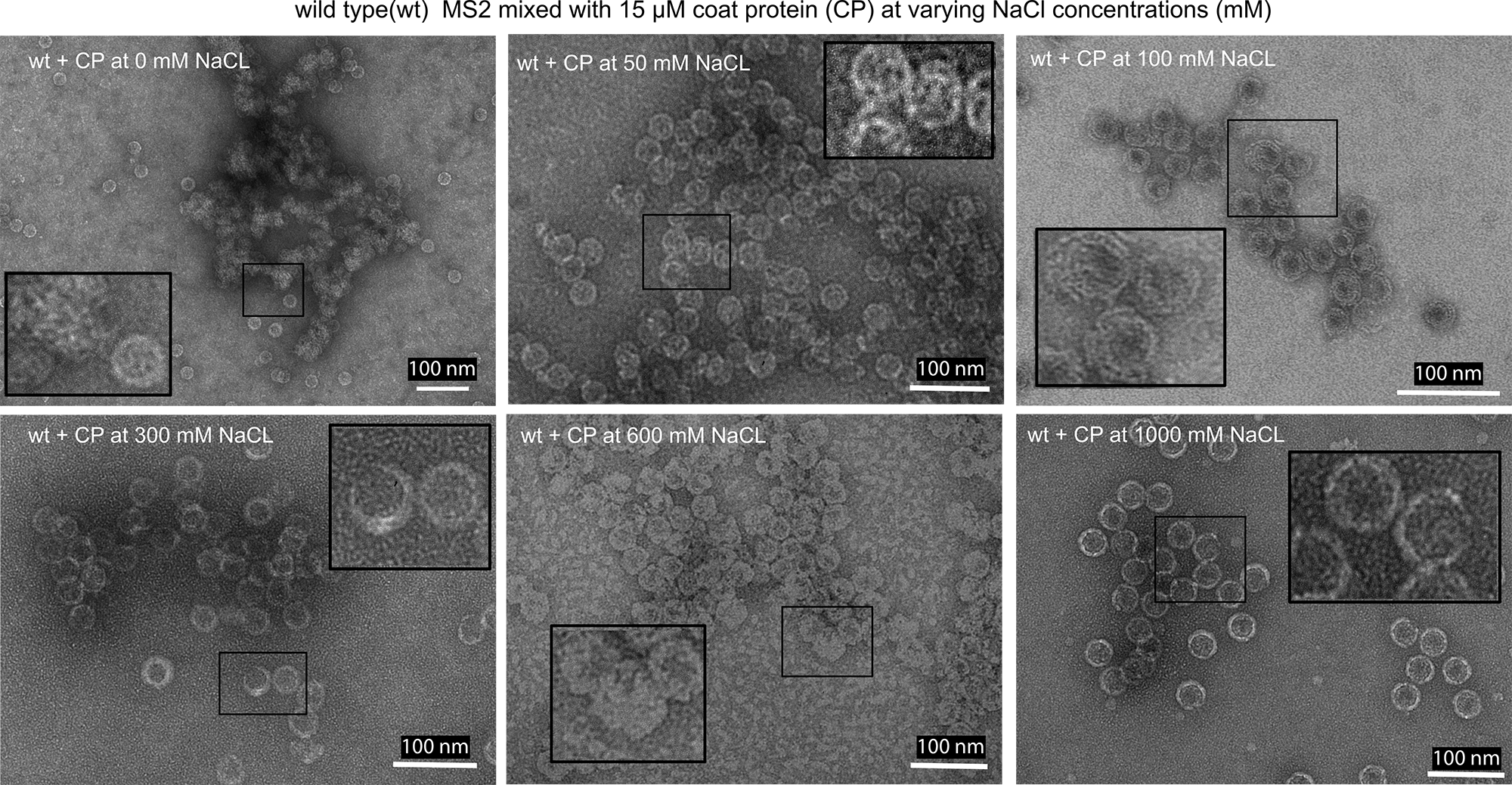}
  \caption{\label{fig:multinacl} TEM images of negatively stained samples of wild-type MS2
    particles mixed with \SI{15}{\micro\molar} coat-protein dimers at
    varying NaCl concentrations.}
\end{figure*}

\begin{figure*}[t]
\centering\includegraphics[width=0.95\textwidth]{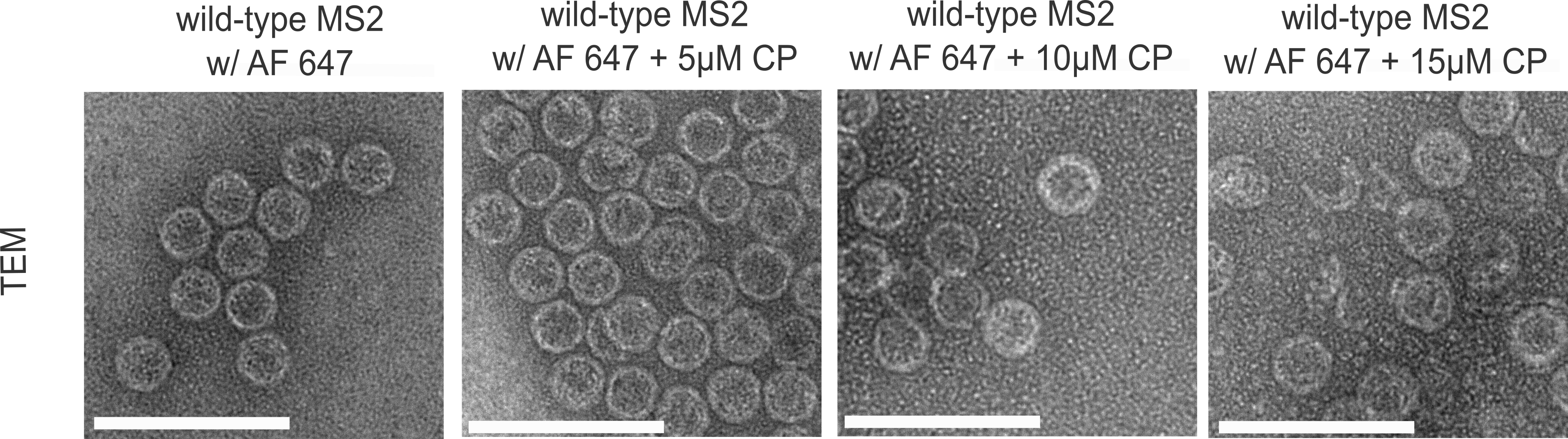}
\caption{\label{fig:multiTEMlabelledwt} Transmission electron microscopy of \SI{50}{\nano\molar}
  wild-type MS2 labeled with Alexafluor 647 (on the RNA) mixed with
  \qtylist{0; 5; 10; 15}{\micro\molar} coat-protein dimers. These images
  show that additional shells are able to form on the labeled particles.
  All scale bars are 100 nm. }
\end{figure*}

\begin{figure*}[t]
  \centering\includegraphics[width=0.95\textwidth]{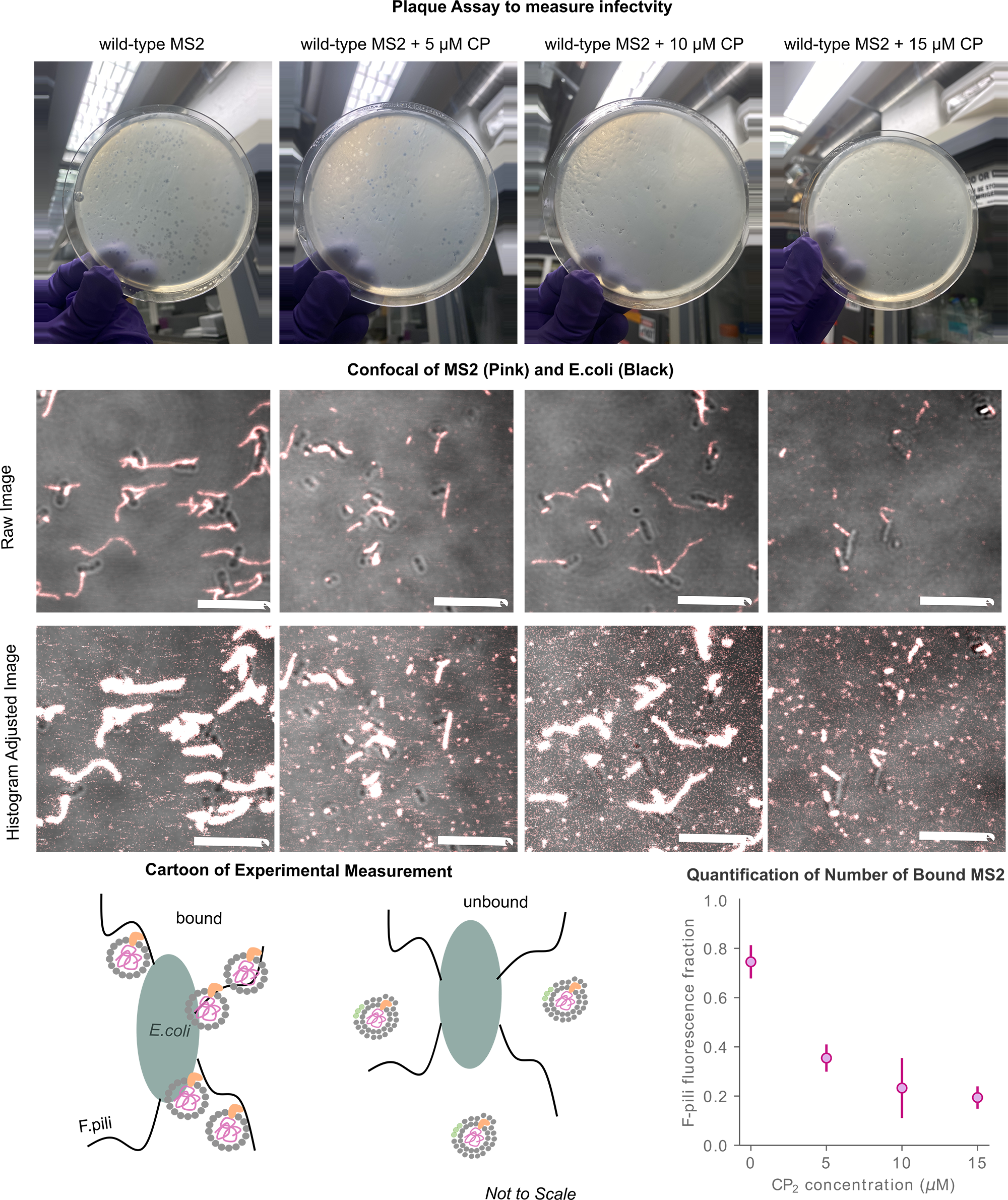}
  \caption{\label{fig:multiinfectivity} Top row shows photographs of
    plaque assays of wild-type MS2 and wild-type MS2 to which coat-protein
    dimers have been added. Middle rows show raw images and -adjusted
    images from confocal microscopy of \textit{E.\@ coli} cells with
    labeled MS2 particles or labeled MS2 particles mixed with 5~$\mu$M,
    10~$\mu$M, and 15~$\mu$M coat-protein dimers. Cartoon at bottom left
    shows how the binding assay selects for particles with unoccluded
    maturation protein. Plot at bottom right shows the F-pili fluorescence
    fraction, a measure of the number of particles bound to F-pili
    relative to the total number of particles).}
\end{figure*}

\end{document}